\documentstyle[12pt,preprint,aps,tighten,epsfig,floats]{revtex}
\def\beq{\begin{equation}}
\def\eeq{\end{equation}}
\def\beqa{\begin{eqnarray}}
\def\eeqa{\end{eqnarray}}
\def\be{\begin{equation}}
\def\ee{\end{equation}}
\def\bea{\begin{eqnarray}}
\def\eea{\end{eqnarray}}
\begin{document}
\title{Phenomenological Issues in the Determination of $\Delta \Gamma_{\rm D}$}
\author{Eugene Golowich$^a$ and Sandip Pakvasa$^b$}
\address{
$^a$Physics Department, University of Massachusetts\\
Amherst, MA 01003\\
$^b$Department of Physics and Astronomy, University of Hawaii at Manoa\\
Honolulu, HI 96822}
\maketitle
\thispagestyle{empty}
\setcounter{page}{0}
\begin{abstract}
We consider the issue of determining the $D^0$-${\bar D}^0$
width difference $\Delta \Gamma_{\rm D}$ experimentally. 
The current situation is reviewed and suggestions for further 
study are given.  We propose a number of $D^0$ decay modes in 
addition to those studied in the recent E791, FOCUS and BELLE lifetime 
determination experiments.  Then we address prospects for 
determining CF - CDS strong phase differences, like $\delta_{K\pi}$ 
which appears in the CLEO study of $D^0 \to K^+ \pi^-$ transitions.  
We show how to extract $\delta_{K^*\pi}$ with CDS data and 
furthermore show when $D \to K_{\rm L}\pi$ data becomes available 
that $\delta_{K\pi}$ can also be obtained.  
\end{abstract}
\section{Introduction}
In the Standard Model, effects of $D^0$-${\bar D}^0$ mixing are 
much smaller than those in the kaon, $B_d$ and 
$B_s$ systems.  However, charm-related experiments of 
increasing sensitivity have been carried out, leading 
to ever-improving bounds on the dimensionless mixing parameters 
$x_{\rm D} \equiv \Delta M_{\rm D}/\Gamma_{\rm D}$ and 
$y_{\rm D} \equiv \Delta \Gamma_{\rm D}/2\Gamma_{\rm D}$.  Most
recently, the E791, CLEO, FOCUS and BELLE collaborations have 
reported on attempts to detect mixing in the 
$D$-meson system.  This has prompted discussion in the literature 
as to whether actual $D$-meson mixing (specifically  
a nonzero $\Delta \Gamma_{\rm D}$) is being seen for the 
first time.~\cite{gang}  Since a rigorous theoretical prediction for 
$\Delta \Gamma_{\rm D}$ is unlikely, experimental progress 
in this area is needed.  In this paper, we discuss specific proposals 
for further work in lifetime difference measurements and 
in experimentally determining the strong phase $\delta$ (which occurs between 
Cabibbo-favored and Cabibbo-doubly-suppressed decays).

\section{Measurements of Lifetime Differences} 
The E791~\cite{e791}, FOCUS~\cite{focus} and BELLE~\cite{belle} 
experiments study the time 
dependence for $D^0 (t) \to K^+ K^-$ (${\rm CP} = +1$ final 
state) and $D^0 (t) \to K \pi$ (CP-mixed final state) under 
the assumption that CP invariance is assumed.  This is reasonable in 
view of both theoretical expectations based on Standard Model 
physics and also recent CLEO results (see Sect.~III). If we adopt 
the convention that ${\cal C}{\cal P} | D^0 \rangle = + | 
{\bar D}^0 \rangle$ and introduce the CP eigenstates 
\beq
|D_{1 \atop 2} \rangle =  {1 \over \sqrt{2}} \left[ 
| D^0 \rangle \pm | {\bar D}^0 \rangle \right] \ \ , 
\label{am2}
\eeq
then $|D_1\rangle$ is CP-even and $|D_2\rangle$ is CP-odd.  
It follows from Eq.~(\ref{am2}) that 
\beq
|D^0 (t) \rangle =  {1 \over \sqrt{2}} \left[ 
| D_1 (t) \rangle + | D_2 (t)  \rangle \right] \ \ ,
\label{am3}
\eeq
where $|D_{1,2}\rangle$ evolve in time with distinct 
masses and decay widths,
\beq
|D_k (t) \rangle = e^{-i M_k t - {1\over 2}\Gamma_k t}
|D_k\rangle \qquad (k = 1,2) \ \ .
\label{am4}
\eeq
If the $K^+K^-$ final state is overlapped with Eq.~(\ref{am3}) 
only the $| D_1 (t) \rangle$ part contributes, leading to 
the exponential decay equation
\beq
\Gamma_{K{\bar K}} (t) = A_{K{\bar K}} e^{-\Gamma_1 t} \ \ .
\label{am5}
\eeq
For the $K \pi$ final state, we express 
the time evolution of $D^0$ as 
\beq
| D^0 (t) \rangle = f_+ (t) | D^0 \rangle + f_- (t) | {\bar D}^0 \rangle 
\label{am6}
\eeq
where 
\beq
f_+ (t) = {1 \over 2} \sum_{k = 1}^2 \ e^{-i M_k t - {1\over 2}\Gamma_k t}
\qquad 
f_- (t) = {1 \over 2} \sum_{i = 1}^2 \ (-)^{i + 1} ~e^{-i M_i t - 
{1\over 2}\Gamma_i t} \ \ .
\label{am7}
\eeq
Then the above conditions $1,2$ imply 
\beq
\Gamma_{K^-\pi^+ + K^+ \pi^-} (t) = A_{K\pi} \left[ e^{-\Gamma_1 t} 
+ e^{-\Gamma_2 t} \right] = 2 A_{K\pi} 
e^{-(\Gamma_1 + \Gamma_2)t/2} 
\cosh\left[(\Gamma_1 - \Gamma_2){t \over 2}\right] \ \ .
\label{am8}
\eeq
The experimental conditions are such that the 
$\cosh$ term in Eq.~(\ref{am8}) is nearly unity. Thus the time 
dependence becomes exponential, allowing determination of 
$(\Gamma_1 + \Gamma_2)/2$.  The E791, FOCUS and BELLE 
experiments measure the quantity $y_{\rm cp}$, 
\beq
y_{\rm cp} \equiv 
\displaystyle{\tau_{D^0 \to K\pi} 
\over \tau_{D^0 \to K^+ K^-}} - 1 = 
\displaystyle{
\Gamma_1 - \Gamma_2 \over 
\Gamma_1 + \Gamma_2} 
\label{intro1}
\eeq
and find 
\beq
y_{\rm cp} = \left\{ 
\begin{array}{ll} 
(0.8 \pm 2.9 \pm 1.0)\% & \qquad {\rm (E791)} \\
(3.42 \pm 1.39 \pm 0.74)\% & \qquad {\rm (FOCUS)} \\ 
(1.0~^{+3.8}_{-3.5}~^{+1.1}_{-2.1})\% & \qquad {\rm (BELLE)} \ \ .
\end{array}
\right.
\label{intro2}
\eeq
Due to its superior sensitivity the FOCUS determination 
dominates, the net result being a positive value for 
$y_{\rm cp}$ of several per cent at about the two standard 
deviation level.

\subsection{Additional Decay Modes}
We urge that additional lifetime studies on CP eigenstates 
of the neutral $D$ be carried out.  It is essential 
to improve the statistical data base and to acquire a 
sensitivity beyond the current $2\sigma$ level. 
Beyond that, there is still no experimental 
input on the pure ${\it CP} = -1$ lifetime.   
By using lifetimes obtained from pure ${\it CP} = \pm 1$ modes, 
one would be determining $\Delta\Gamma$ directly rather than 
comparing an average of ${\it CP} = \pm 1$ lifetimes with that 
of ${\it CP} = + 1$.   

There are a number of opportunities for further study, 
each final state occurring in $D^0$ decay being a potential candidate. 
We shall discuss just a limited number of these 
in the following, citing disadvantages as well as advantages.  
An important subset of our list has modes which contain a 
pair of mesons, each of which is self-conjugate under the CP
operation.  If each member of the pair has spin-zero, the orbital 
angular momentum is S-wave and the CP value of the two-particle state 
is simply the product of the individual CP values.  
If one meson has spin-zero and the other has spin-one, then 
conservation of angular momentum requires the particles 
to be in a P-wave.  In this case, the CP value of the 
two-particle state becomes {\it minus} the product of the 
individual CP values.  

Vertex identification is a key to 
a successful $D^0$ lifetime measurement.  Starting from 
its (assumed known) production point, the 
$D^0$ will travel unobserved and ultimately decay into some 
final state particles.  In the best case, all these 
are charged and the decay point becomes well determined.  
In the worst case, each primary decay product is neutral 
and, if unstable, decays itself into neutral particles.  
Then even with calorimetric information, attempting to fix 
a decay point is problematic.  For reference, in the E791, 
FOCUS and BELLE experiments the detected modes 
($K\pi$ and $K {\bar K}$) had just two particles (both 
charged) in each final state and the branching fractions 
were ${\cal B}_{K^-\pi^+}= (3.83 \pm 0.09)\cdot 10^{-2}$ and   
${\cal B}_{K^-K^+}= (4.25 \pm 0.16)\cdot 10^{-3}$.~\cite{pdg}  

Many $D^0$ decay modes contain neutral kaons in 
the final state.  The neutral kaons will in turn decay 
as $K_{\rm S}$ or $K_{\rm L}$ mesons.  For a lifetime 
determination measurement, a $K_{\rm S}$ mode is superior 
to a $K_{\rm L}$ mode because: (i) the $K_{\rm S}$ 
detection efficiency is rather larger than the $K_{\rm L}$ 
detection efficiency, so the statistics will be better for the 
former, (ii) the $K_{\rm L}$ decay occurs further from 
the $D^0$ decay vertex, so its background problem is more severe.  
Both of these considerations are inherent for any detector.  
However, since progress in dealing with $K_{\rm L}$ 
detection is anticipated the 
$K_{\rm L}$ modes should not be totally disregarded.  
To summarize, $K_{\rm S}$ detection is 
easier and can be done now whereas $K_{\rm L}$ is harder and 
may be done later, although not as well.  Finally, we note 
that PDG listings give branching fractions for 
$D \to {\bar K}^0 X$ ($X$ denotes other final state particles) 
rather than for $D \to K_{\rm S,L}X$.  It will suffice below 
to use the approximations 
\beq
\Gamma_{D^0 \to K_{\rm S}X} 
\simeq \Gamma_{D^0 \to K_{\rm L}X} \simeq 
{1 \over 2} \Gamma_{D^0 \to {\bar K}^0X} \ \ .
\label{int}
\eeq
These relations are not exact because 
decay into $K_{\rm S}$ or $K_{\rm L}$ is subject to 
interference between Cabibbo favored (CF) and 
Cabibbo doubly suppressed (CDS) modes.~\cite{by}  We discuss 
aspects of this interference in the next section.  

Now we turn to the list of additional possible modes, 
partitioned according to the CP of the final states 
and presented as ${\rm CP} = -1$, ${\rm CP} = +1$ and 
${\rm CP}$-mixed. 

\begin{center}
{\it Pure ${\it CP} = - 1$ Modes} 
\end{center}

\begin{enumerate}
\item $K_{\rm S} \phi$: Both the $K_{\rm S}$ and 
$\phi$ decay into charged final states, so this mode is an 
attractive one as regards particle detection.  Since the 
$\phi\to K^+K^-$ transition is a strong decay, it occurs right 
at the $D^0$ decay vertex.  Also, the $\phi$ has a 
narrow decay width. The branching fraction for this mode is 
acceptably large (${\cal B}_{\phi K_{\rm S}}= 
(4.3 \pm 0.5)\cdot 10^{-3}$). 
\item $K_{\rm S} \omega$: Although the branching fraction is 
respectable (${\cal B}_{K_{\rm S}\omega}= 
(1.05 \pm 0.2)\cdot 10^{-2}$), the $\omega$ decays 
predominantly via the three-body mode $\pi^+\pi^-\pi^0$ which 
renders it more difficult regarding identification of the decay vertex.  
\item $K_{\rm S} \rho^0$: In this case, the 
branching fraction is not unattractive (${\cal B}_{K_{\rm S}\rho^0}= 
(0.61 \pm 0.09)\cdot 10^{-2}$), and the $K_{\rm S} \rho^0$ 
final state would decay into all charged particles.  However, 
the larger width of the $\rho^0$ (compared to the $\phi$) makes 
detection relatively more difficult.  
\item $K_{\rm S} \pi^0$: This mode has a reasonably large 
branching fraction ${\cal B}_{K_{\rm S} \pi^0} = 
(1.06 \pm 0.11)\cdot 10^{-2}$).  However, the presence of the 
$\pi^0$ hinders accurate vertex identification.
\item $K_{\rm S}\eta$ and $K_{\rm S}\eta'$: Both these modes 
are potentially interesting since the branching fractions are not 
highly suppressed $\left( {\cal B}_{K_{\rm S} \eta} = (3.5 \pm 
0.5)\cdot 10^{-3} \right.$  and $\left. {\cal B}_{K_{\rm S} \eta'} 
= (8.5 \pm 1.3)\cdot 10^{-3} \right)$.  The problem of vertex 
ID for a final state $\eta$ and $\eta'$ would resemble 
that for a final state $\omega$. 
\end{enumerate}  

\begin{center}
{\it Pure ${\it CP} = +1$ Modes} 
\end{center}

\begin{enumerate}
\item $\pi^+\pi^-$: This mode provides a clean ${\rm CP} = +1$ 
signal but has the disadvantage of a small branching fraction 
(${\cal B}_{\pi^-\pi^+}= (1.52 \pm 0.09)\cdot 10^{-3}$), about 
three times less than $K^+K^-$.  Also backgrounds could be a 
problem since since there are more $\pi^+\pi^-$ combinations 
in a typical event (although in the $D^0$ rest frame, the two 
pions emerge back-to-back with larger momenta than any other 
final state).  
\item $K_{\rm L} \eta$ and $K_{\rm L} \eta'$: Although the 
branching fractions equal those for $K_{\rm S}\eta$ and $K_{\rm
S}\eta'$, detection of the $K_{\rm L}$ presents difficulties, as 
discussed earlier.  
\item $\pi^0 \phi$: Particle ID is more of an issue than for 
the $K_{\rm S} \phi$ mode as the neutral $\pi^0$ decays via 
the chargeless two-photon mode.  Even though the 
branching fraction here is comparatively large 
(${\cal B}_{\pi^0 K_{\rm S}}= (1.05 \pm 0.11)\cdot 10^{-2}$), 
it is not sufficient to compensate for the detection problem.  
Moreover, other decay modes containing $\pi^0$'s 
would be a source of background.  
\item $K_{\rm S} f_0 (980)$: This mode consists of a 
scalar-pseudoscalar pair in an S-wave, and has 
${\rm CP} = + 1$ for $K_{\rm S}$ and  ${\rm CP} = + 1$ for 
$f_0 (980)$.  Although the $K_{\rm S}$ and $f_0 (980)$ each decay 
into charged particles, the branching ratio is small 
(${\cal B}_{K_{\rm S}f_0(980)}= (2.9 \pm 0.8)\cdot 10^{-3}$).  
Similar comments apply to the $K_{\rm S} f_0 (1370)$ final 
state and to the  $K_{\rm S} f_2 (1270)$ (except that here 
the final state is D-wave).
\item $\phi\rho^0$:  This mode will have positive CP 
provided the $\phi$ and $\rho^0$ are in an S-wave of D-wave state.  
Both decay strongly into charged particles, so the decay point 
will have four emergent tracks.  The branching fraction is rather 
small $\left( {\cal B}_{\phi\rho^0} = (6 \pm 3) 
\cdot 10^{-4} \right).$  
\item $K_{\rm L}\pi^0$: This final state has the same branching fraction 
as $K_{\rm S}\pi^0$, but an even greater detection problem 
due to the $K_{\rm L}$.  In practical terms, vertex identification 
would be an insurmountable obstacle. 
\end{enumerate}

\begin{center}
{\it Mixed ${\it CP} = \pm 1$ Modes} 
\end{center}
For definiteness consider a Dalitz plot analysis for the neutral 
($Q = 0$) three-body state $(\pi\pi {\bar K})_{\rm Q = 0}$.  
There will be resonance bands corresponding to the quasi 
two-body modes $(\rho {\bar K})_{\rm Q = 0}$ and $(\pi 
{\bar K}^{*})_{\rm Q = 0}$. Although neither the $\rho$ nor the 
${\bar K}^*$ is a particularly narrow resonance, these decays 
are CKM dominant so the branching ratios are relatively large.  
Specific examples of mixed ${\it CP} = \pm 1$ modes are:
\begin{enumerate}
\item $\pi^+ {\bar K}^{*-}$: This quasi two-particle 
state has a large branching fraction 
${\cal B}_{\pi^+ {\bar K}^{*-}}= (5.0 \pm 0.4)\cdot 10^{-2}$ 
and there are the two measureable ${\bar K}^*$ decay 
modes $K^{*-}\to \pi^0 K^-$ and $K^{*-}\to \pi^- {\bar K}^0$. 
The latter provides a rather clean three-body configuration, 
$\pi^+(\pi^-K_{\rm S})$ where the parentheses stress the 
$K^{*-}$ parentage. 
\item $\rho^+ K^{-}$: The largest branching fraction among 
all quasi two-body final states for $D^0$ decay occurs here, 
${\cal B}_{\rho^+ K^{-}}= (10.8 \pm 0.9)\cdot 10^{-2}$.  
The $\rho^+$ decay proceeds through only the mode 
$\rho^+\to \pi^+\pi^0$. 
\item $\pi^0 {\bar K}^{*0}$: The branching fraction 
${\cal B}_{\pi^0 {\bar K}^{*0}}= (3.1 \pm 0.4)\cdot 10^{-2}$ 
is relatively large.  The associated three-body configurations 
will be $\pi^0 (\pi^-{\bar K}^{0})$ and the less useful 
$\pi^0 (\pi^0 K^{-})$. 
\end{enumerate}  

\section{Measurements of Wrong-sign $D^0$ Transitions}
Another study which impacts on determining $\Delta\Gamma_{\rm D}$ is 
the CLEO experiment~\cite{cleo} which studies the decay rate for 
$D^0 (t) \to K^+ \pi^-$. This wrong-sign process can 
be produced both indirectly, from mixing followed by a CF decay, 
and directly, from CDS decay.  
The decay rate is given in the CP-invariant limit by 
\beq
r(t) = e^{- t} \left[ R_{\rm D} + \sqrt{R_{\rm D}}~ y' \cdot t 
+ R_{\rm M} \cdot t^2 \right] \ \ .
\label{intro3}
\eeq
The $R_{\rm D}$ term arises from CDS decay, the $R_{\rm M}$ 
term from mixing and the $\sqrt{R_{\rm D}}$ term from 
interference between the two.  We also have the definitions 
\beq
y' \equiv y \cos\delta - x \sin\delta \ , \qquad 
x' \equiv x \cos\delta + y \sin\delta \ \ . 
\label{intro4}
\eeq
The parameter $\delta$ is the (strong-interaction) phase difference 
between between the CF and CDS amplitudes,
\beq
\delta \equiv \delta_{K\pi}^{\rm (ch)} 
\equiv \delta_{-+} - \delta_{+-} \ \ ,
\label{intro4a}
\eeq
where the phases $\delta_{-+}$ and $\delta_{+-}$ 
appear in the amplitudes 
\beq
{\cal M}_{D^0 \to K^- \pi^+} = |{\cal M}_{D^0 
\to K^- \pi^+}| ~e^{i \delta_{-+}} \qquad 
{\cal M}_{D^0 \to K^+ \pi^-} = |{\cal M}_{D^0 
\to K^+ \pi^-}| ~e^{i \delta_{+-}} \ \ .
\label{intro4b}
\eeq
Note that we sharpen the notation for $\delta$ ($\delta \to 
\delta_{K\pi}^{\rm (ch)}$) in Eq.~(\ref{intro4a}) because we 
will encounter several analogous phases in our analysis. 

The CP-invariant 
rate formula of Eq.~(\ref{intro3}) can be generalized to 
incorporate various sources of CP-violation (CPV),~\cite{nir}
\beqa
R_{\rm D} &\to & R_{\rm D} \left( 1 + A_{\rm D} \right) 
\qquad {\rm (CDS)} \ \ , \nonumber \\
y' &\to& y' \left( 1 + A_M/2 \right) \qquad {\rm (mixing)} \ \ , 
\label{intro5} \\
\delta_{K\pi}^{\rm (ch)} 
&\to& \delta_{K\pi}^{\rm (ch)} + \phi \qquad 
{\rm (interference)} \ \ , \nonumber 
\eeqa
where $A_{\rm D}$, $A_M$ and $\phi$ parameterize the extent of 
CP violation.  When the data is fit to include the 
effects of CP violation none is found, 
\beq
A_{\rm M} = 0.23^{+.63}_{-.80} \pm .01 \ , \qquad 
A_{\rm D} = -0.01^{+.16}_{-.17} \pm .01 \ , \qquad 
\sin\phi = 0.00 \pm .60 \pm .01 \ \ .
\label{intro6}
\eeq
In the same fit one finds at $95\%$ D.L. 
\beq
x' = (0 \pm 1.5 \pm 0.2)\% \qquad {\rm and} \qquad 
y'= (-2.5^{+1.4}_{-1.6} \pm 0.3) \% 
\label{intro7}
\eeq
or equivalently 
\beq
|x'| < 2.9 \% \qquad {\rm and} \qquad 
-5.8 \% < y' < 1.0 \% \ \ .
\label{intro8}
\eeq

Given the present strength of the CLEO signals for $x'$ and $y'$, 
it is prudent to cite the results as bounds as in Eq.~(\ref{intro8}).
One expects future experiments to reduce the statistical and 
systematic uncertainties.  Even so, ignorance of the phase 
$\delta_{K\pi}^{\rm (ch)}$ will hamper efforts to compare the 
FOCUS/E791/BELLE results with those from CLEO. 

\subsection{On the Determination of $y'$}
Can theory alone provide the value of $\delta_{K\pi}^{\rm (ch)}$?  
Symmetry considerations are of only limited use.  
It is known that $\delta_{K\pi}^{\rm (ch)}$ 
vanishes in the SU(3) invariant world~\cite{ktwz,hk}, and 
this result has been 
recognized~\cite{lw} in discussing aspects of the wrong-sign $D^0$ 
transitions.   Thus, calculating the value 
of $\delta_{K\pi}^{\rm (ch)}$ necessarily involves the physics of SU(3) 
breaking.  Unfortunately, our limited understanding of physics in the 
charm region (especially the complicating effects of QCD) 
makes it difficult to perform reliable calculations.~\cite{hn}  
It is, perhaps, not too surprising to find rather different statements
in the literature about $\delta_{K\pi}^{\rm (ch)}$ 
depending on the underlying 
approach.  In one analysis~\cite{bp}, the findings of 
Ref.~\cite{cc} and Ref.~\cite{blp} are shown to imply rather 
small values for $\delta_{K\pi}^{\rm (ch)}$, less than $15^o$. 
However, the resonance model of Ref.~\cite{fnp} has 
considerably greater SU(3) breaking and obtains values 
as large as $\delta_{K\pi}^{\rm (ch)} \sim 30^o$.  The largest 
value cited for $\delta_{K\pi}^{\rm (ch)}$ appears in 
Ref.~\cite{gang} which shows that accepting the central 
values of the FOCUS and CLEO experiments leads to 
$\delta_{K\pi}^{\rm (ch)}$ in the second quadrant.  However, 
it has been argued~\cite{gr} that within a reasonable range of 
SU(3)-breaking parameters it is not possible to arrive at very large 
values of $\delta_{K\pi}^{\rm (ch)}$ ($45^o$ or larger) of the 
type considered in Ref.~\cite{gang}. 

In view of this state of affairs, it makes sense to explore what 
experiment can teach us.

\subsection{Doing without $K_{\rm L}$ Data} 
Recalling our comments in the previous section on 
the relative measureability of $K_{\rm S,L}$ modes, 
we begin by assuming that only a data set {\it not} containing 
final state $K_{\rm L}$'s is available.  
The inclusion of $K_{\rm L}$ data is covered later.

Our first conclusion concerns the CDS $D \to K \pi$ decays: 
$D^0 \to K^+ \pi^-$, $D^+ \to K^+ \pi^0$, 
$D^0 \to K^0 \pi^0$ and $D^+ \to K^0 \pi^+$. At present, only 
the first of these has been observed (${\cal B}_{K^-\pi^+} = 
(1.46 \pm 0.30)\cdot 10^{-4}$).  If only $K_{\rm S}$ data 
is used, then neither $D^0 \to K^0\pi^0$ nor 
$D^+ \to K^0\pi^+$ modes can be determined experimentally. 
This can be understood by considering CF and CDS 
transitions having a neutral kaon in the final state: 
\begin{eqnarray*}
\begin{array}{c||c}
\multicolumn{2}{c}{D \to {\bar K}^0 \pi \ {\rm CF\ Decays}} \\ \hline 
{\rm Transition} & {\rm Final\ State} \\ \hline 
D^0 \to {\bar K}^0 \pi^0 &  K_{\rm S} \pi^0 \\
D^+ \to {\bar K}^0 \pi^+ &  K_{\rm S} \pi^+ \\ \hline 
\end{array}
\qquad 
\begin{array}{c||c}
\multicolumn{2}{c}{D \to K^0 \pi \ {\rm CDS\ Decays}}  \\ \hline
{\rm Transition} & {\rm Final\ State} \\ \hline 
D^0 \to K^0 \pi^0 &  K_{\rm S} \pi^0 \\
D^+ \to K^0 \pi^+ &  K_{\rm S} \pi^+ \\ \hline
\end{array}
\end{eqnarray*} 
For both CDS transitions $D^0 \to K^0\pi^0$ and $D^+ \to K^0\pi^+$, 
the $K^0$ will decay via the same $K_{\rm S} \to \pi^+\pi^-$ 
mode as the CF transitions $D^0 \to {\bar K}^0\pi^0$ and $D^+ \to 
{\bar K}^0\pi^+$.  Since the CF decays dominate, extracting 
information about CDS final states containing a $K^0$ 
from just $K_{\rm S}$ detection will be impossible.  This negates 
performing a direct experimental measurement of $\delta_{K\pi}^{\rm (ch)}$. 

What is the situation for other possible final states like $K \rho$ 
or $K^* \pi$?  Clearly, the same no-go result will hold for the 
$D \to K \rho$ decays.   This leaves only the case of 
$D \to K^* \pi$.  Since the $K^*$ decays strongly into two 
different charge combinations of $K \pi$, each $D \to K^* \pi$  
transition will have two final configurations.  Continuing to assume 
that only the $K_{\rm S}$ mode in $K^0$ and ${\bar K}^0$ is 
observed, we obtain the following list: 
\begin{eqnarray*}
\begin{array}{c||c|c}
\multicolumn{3}{c}{D \to {\bar K}^* \pi \ {\rm CF\ Decays}} \\ \hline 
{\rm Transition} & {\rm FS1} &  {\rm FS2} 
\\ \hline 
D^0 \to K^{*-} \pi^+ & (K^- \pi^0) \pi^+ & (K_{\rm S} \pi^-)\pi^+ \\
D^0 \to {\bar K}^{*0} \pi^0 & (K^- \pi^+) \pi^0 & (K_{\rm S} \pi^0)\pi^0
\\
D^+ \to {\bar K}^{*0} \pi^+ & (K^- \pi^+) \pi^+ & (K_{\rm S} \pi^0)\pi^+ 
\\ \hline 
\end{array}
\quad 
\begin{array}{c||c|c}
\multicolumn{3}{c}{D \to K^* \pi \ {\rm CDS\ Decays}}  \\ \hline
{\rm Transition} & {\rm FS1} &  {\rm FS2} 
\\ \hline 
D^0 \to K^{*+} \pi^- & (K^+ \pi^0) \pi^- & (K_{\rm S} \pi^+)\pi^- \\
D^0 \to K^{*0} \pi^0 & (K^+ \pi^-) \pi^0 & (K_{\rm S} \pi^0) \pi^0 \\
D^+ \to K^{*+} \pi^0 & (K^+ \pi^0) \pi^0 & (K_{\rm S} \pi^+) \pi^0 \\
D^+ \to K^{*0} \pi^+ & (K^+ \pi^-) \pi^+ & (K_{\rm S} \pi^0) \pi^+ 
\\ \hline
\end{array}
\end{eqnarray*}
Each $K\pi$ arising from $K^*$ decay is enclosed in parentheses 
and $FS1$, $FS2$ are the two three-body final states 
per $D$ decay.  Each CDS transition with a $K^{*0}$ 
in the final state has a configuration (FS2) identical to that 
of a CF transition with a ${\bar K}^{*0}$ in the final state. 
However, the other configurations (FS1) each contain a charged 
kaon and thus distinguish between CF and CDS decays.  

Thus, all four $D \to K^* \pi$ 
CDS decays {\it can} be utilized.  In those final states containing 
a $K^{*+}$, both configurations FS1 and FS2 will have a unique 
signature (it is, however, necessary to employ a Dalitz 
plot analysis to properly identify which `$K\pi$' composite is a 
product of $K^*$ decay).  For final states with a $K^{*0}$, 
there will be a reduction factor of $2/3$ in the number of events 
since only the configuration FS1 can be used, {\it i.e.}
\beq  
\Gamma_{D^0 \to (K^+\pi^-)\pi^0} = {2\over 3} ~
\Gamma_{D^0 \to K^{*0}\pi^0} \qquad {\rm and} \qquad 
\Gamma_{D^+ \to (K^+\pi^-)\pi^+} = {2\over 3} ~
\Gamma_{D^+ \to K^{*0}\pi^+} \ \ .
\label{ph1}
\eeq

Thus we are led to analyze the phenomenology of $D \to K^* \pi$
transitions in both the CF and CDS sectors. 
\subsubsection{Cabibbo Favored (${\bar K}^*\pi$) Decays}
There are three Cabibbo favored (CF) $D \to {\bar K}^* \pi$ 
decays,
\beq
D^0 \to K^{*-} \pi^+\ , \qquad D^0 \to {\bar K}^{*0} \pi^0\ , \qquad 
D^+ \to {\bar K}^{*0} \pi^+\ \ .
\label{cf1}
\eeq
These proceed through the QCD-corrected $\Delta S = \Delta C = \pm 1$ 
weak hamiltonian, which takes the form~\cite{buras} 
\beq
{\cal H}_{\rm W}^{\rm (CF)} = c_- {\cal H}_-^{\rm (CF)} + 
c_+ {\cal H}_+^{\rm (CF)} \ \ .
\label{cf1a}
\eeq
${\cal H}_-^{\rm (CF)}$ and ${\cal H}_+^{\rm (CF)}$ 
transform alike under isospin, as the $I_3 = +1$ member 
of an isotriplet. Under SU(3), however, 
${\cal H}_-^{\rm (CF)}$ belongs to 
${\bf 6}\oplus{\bf {6^*}}$ and ${\cal H}_+^{\rm (CF)}$
to ${\bf {15}}\oplus{\bf {15^*}}$.~\cite{dh,cam} 
The coefficients $c_\pm$ encode the short distance, perturbative 
part of the QCD corrections. At energy 
scale $M_{\rm W}$, $c_\pm$ have essentially equal 
magnitudes.  As the energy scale is lowered, the coefficient $c_-$ is 
enhanced whereas $c_+$ is suppressed.  

Using just the isospin property of ${\cal H}_{\rm W}^{\rm (CF)}$, we 
express the above decay amplitudes as\footnote{Equivalent 
formulae can be written for ${\bar K} \pi$ and ${\bar K}\rho$.}  
\beqa
{\cal M}_{K^{*-}\pi^+} &=& A_1 e^{i \delta_1} 
+ {1\over 2} A_3 e^{i \delta_3} \ \ , \nonumber \\
{\cal M}_{{\bar K}^{*0}\pi^0} &=& -{1 \over \sqrt{2}} A_1 e^{i \delta_1} 
+ {1\over \sqrt{2}} A_3 e^{i \delta_3} \ \ , \label{cf2} \\
{\cal M}_{{\bar K}^{*0}\pi^+} &=& {3 \over 2} A_3 e^{i \delta_3} 
\ \ , \nonumber 
\eeqa
where the subscripts represent twice the isospin of the final state 
${\bar K}^*\pi$ composites.  Observe that these amplitudes obey 
the sextet-dominance constraints ${\cal M}_{{\bar K}^{*0}\pi^+} 
= 0$ and ${\cal M}_{K^{*-}\pi^+} = - \sqrt{2} 
{\cal M}_{{\bar K}^{*0}\pi^0}$.   Upon either expanding the 
relation $0 = \langle {\bar K}^{*0} \pi^+ | [ I_+ , 
{\cal H}_{\rm W}^{\rm (CF)}] | D^0 \rangle$ or utilizing the 
amplitude relations of Eq.~(\ref{cf2}), one arrives at the 
isospin sum rule~\cite{bhp,eg,jr,mg}
\beq
{\cal M}_{K^{*-}\pi^+}  + \sqrt{2} {\cal M}_{{\bar K}^{*0}\pi^0} 
- {\cal M}_{{\bar K}^{*0}\pi^+} = 0 \ \ .
\label{cf3}
\eeq
Of interest to us here are the phase difference and amplitude ratio, 
\beqa
\cos\left( \delta_1 - \delta_3 \right) &=&
{ \sqrt{2} \over 4} \cdot { 3 \Gamma_{K^{*-}\pi^+} 
+ \Gamma_{{\bar K}^{*0}\pi^+} 
- 6 \Gamma_{{\bar K}^{*0}\pi^0} \over \left[ 
\Gamma_{{\bar K}^{*0}\pi^+} 
\left( 3 \Gamma_{K^{*-}\pi^+} + 3 \Gamma_{{\bar K}^{*0}\pi^0} 
- \Gamma_{{\bar K}^{*0}\pi^+} \right) \right]^{1/2}}
\ \ , \nonumber \\
{A_3 \over A_1} &=& \left[ { 2 \Gamma_{{\bar K}^{*0}\pi^+} \over 
3 \Gamma_{K^{*-}\pi^+} + 3 \Gamma_{{\bar K}^{*0}\pi^0} 
- \Gamma_{{\bar K}^{*0}\pi^+} } \right]^{1/2} \ \ .
\label{cf4}
\eeqa
The most recent data compilation~\cite{pdg} gives for the 
three CF modes $D \to {\bar K}^*\pi, {\bar K} \pi, {\bar K}\rho$, 
\begin{eqnarray*}
\begin{array}{c||c|c}
{\rm Mode} & \delta_1 - \delta_3 & A_3 / A_1 \\ \hline 
{{\bar K}^*} \pi &  {103.9^o}^{+ 17.2^o}_{-17.8^o} & 
0.25^{+0.02}_{-0.01} \\
{\bar K} \pi &  {90.2^o}^{+ 7.1^o}_{-8.2^o} & 0.37 \pm 0.03 \\
{\bar K} \rho &  0.0^o \pm 44.9^o & 0.39 \pm 0.10 \\ \hline
\end{array}
\end{eqnarray*}

The preceding equations can of course be used to obtain phase 
relations in addition to those in the above table, 
{\it e.g.} for the ${\bar K}\pi$ system, 
\beq
\delta_{K^-\pi^+} - \delta_3 = 79.5^o \ , \qquad 
\delta_{{\bar K}^0\pi^0} - \delta_3 = 110.3^o \ \ , 
\label{add}
\eeq
and so on.  Staying temporarily with the $D \to {\bar K} \pi$ 
decays, let us 
compare physics of the real world with that of a world which is 
SU(3) symmetric and in which $c_+ = 0$:   
\begin{eqnarray*}
\begin{array}{r|c|c}
 & \Gamma_{D^0\to {\bar K}^{0}\pi^0} /
\Gamma_{D^0\to K^{-}\pi^+}  & 
\Gamma_{D^+\to {\bar K}^{0}\pi^+} /
\Gamma_{D^0\to K^{-}\pi^+} 
\\ \hline 
{\rm Hypothetical\ world} & 0.5 & 
0.0 \\ 
{\rm Real\ world} & 0.551 \pm 0.006 & 
0.296 \pm 0.028
\end{array}
\end{eqnarray*}
For these rates, at least, 
the agreement between the real world and the hypothetical world
is not unreasonable.  That the hypothetical world is, in some sense, 
nearby the real world will be useful later as a guiding principle in our 
study of the CDS amplitudes.  This comparison between the real world 
and the hypothetical SU(3) world having $c_+ = 0$ explains 
the small observed values of $A_3 / A_1$.  In an SU(3) symmetric 
world, the limit $c_+ = 0$ would correspond to $A_3 / A_1 = 0$ 
for the $D \to {\bar K}\pi$ amplitudes.  
Although the precise values of $c_\pm$ 
depend on the renormalization scheme (involving both the choice 
of operator basis and of renormalization scale $\mu$), a typical 
numerical value is $c_+ /c_- \simeq 0.5$ 
for the range $2.0 \ge \mu ({\rm GeV}) \ge 1.5$.~\cite{buras}  
The short distance effects embodied in $c_\pm$ 
account for much of the suppression for $A_3/A_1$ observed in the
above table, the rest arising from the operator matrix elements.   
Operator matrix elements play a much larger role in the kaon 
system ($\Delta I = 1/2$ rule) where QCD effects are more 
powerful.  

\subsubsection{Cabibbo Doubly Suppressed ($K^*\pi$) Decays}
Corresponding to the three $D \to {\bar K}^* \pi$ CF decays of 
Eq.~(\ref{cf1}) are the following {\it four} $D \to K^*\pi$ decays 
in the Cabibbo doubly suppressed (CDS) sector, 
\beq
D^0 \to K^{*+} \pi^-\ , \qquad D^+ \to K^{*0} \pi^+\ , \qquad 
D^0 \to K^{*0} \pi^0 \ , \qquad D^+ \to K^{*+} \pi^0 \ \ .
\label{cds1}
\eeq
The CDS weak hamiltonian has $\Delta S = - \Delta C = \pm 1$ 
and is written analogous to Eq.~(\ref{cf1a}), 
\beq
{\cal H}_{\rm W}^{\rm (CDS)} = c_- {\cal H}_-^{\rm (CDS)} + 
c_+ {\cal H}_+^{\rm (CDS)} \ \ , 
\label{cds3}
\eeq
but now ${\cal H}_-^{\rm (CDS)}$ and ${\cal H}_+^{\rm (CDS)}$ 
behave differently under isospin, transforming 
respectively as an isosinglet 
and as the $I_3 = 0$ member of an isotriplet.  
Under SU(3), ${\cal H}_-^{\rm (CDS)}$ (like 
${\cal H}_-^{\rm (CF)}$) transforms as a member of 
${\bf 6}\oplus{\bf {6^*}}$ and ${\cal H}_+^{\rm (CDS)}$ 
(like ${\cal H}_+^{\rm (CF)}$) transforms as a member of 
${\bf {15}}\oplus{\bf {{15}^*}}$. 

Performing isospin decompositions of the 
CDS decay amplitudes yields 
\beqa
{\cal M}_{K^{*+}\pi^-} &=& \sqrt{2} {\bar A}_a e^{i {\bar \delta}_1} 
- \sqrt{2} {\bar A}_3 e^{i {\bar \delta}_3} \ , \qquad 
{\cal M}_{K^{*0}\pi^+} = \sqrt{2} {\bar A}_b e^{i {\bar \delta}_1} 
+ \sqrt{2} {\bar A}_3 e^{i {\bar \delta}_3} \ , \nonumber \\
{\cal M}_{K^{*0}\pi^0} &=& - {\bar A}_a  e^{i {\bar \delta}_1} - 2 
{\bar A}_3 e^{i {\bar \delta}_3} \ , \qquad \phantom{xxx}
{\cal M}_{K^{*+}\pi^0} = - {\bar A}_b e^{i {\bar \delta}_1} + 
2 {\bar A}_3 e^{i {\bar \delta}_3} \ \ , 
\label{cds2} 
\eeqa
where the CDS isospin moduli and phases are labelled with 
super-bars.  Corresponding to the above four decay CDS decay 
amplitudes are the four physical observables ${\bar A}_a$, 
${\bar A}_b$, $A_3$ and ${\bar \delta}_1 - {\bar \delta}_3$. 
Two distinct $I=1/2$ moduli (${\bar A}_a$ and ${\bar A}_b$) occur 
because there are two independent sources of the $I = 1/2$ final state, 
the isoscalar ${\cal H}_-^{\rm (CDS)}$ and the isovector 
${\cal H}_+^{\rm (CDS)}$,  
\beq
{\bar A}_a \equiv \sqrt{3} A_1^{(-)} + A_1^{(+)} \qquad {\rm and}
\qquad 
{\bar A}_b \equiv \sqrt{3} A_1^{(-)} - A_1^{(+)} \ \ .
\label{cds2a}
\eeq

We will need to determine the phase ${\bar \Delta}$  
\beq
{\bar \Delta} \ \equiv \ {\bar \delta}_1 - {\bar \delta}_3
\label{ph3}
\eeq
and the moduli ratios $r$, $R$, 
\beq
r \equiv {{\bar A}_3 \over {\bar A}_a} 
\qquad {\rm and} \qquad R \equiv {{\bar A}_b \over {\bar A}_a} \ .
\label{ph4}
\eeq
Since ${\bar A}_a$, ${\bar A}_b$ and ${\bar A}_3$ are moduli, 
we have $r > 0$ and $R > 0$.
In addition, we note that ${\bar A}_3/{\bar A}_b = r/R$.  

Taking the absolute square of each relation in 
Eq.~(\ref{cds2}) and forming ratios gives 
\beqa
{\cal R}_1 &=& {2 + 2 r^2 - 4 r \cos\bar\Delta \over
1 + 4 r^2 + 4 r \cos\bar\Delta} \ , \qquad 
{\cal R}_2 = {2 + 2 r^2 - 4 r \cos\bar\Delta \over
2 R^2 + 2 r^2 + 4 r R \cos\bar\Delta} \ , \nonumber \\
{\cal R}_3 &=& {2 + 2 r^2 - 4 r \cos\bar\Delta \over
R^2 + 4 r^2 - 4 r R \cos\bar\Delta} \ \ , \label{a1}
\eeqa
where the $\{ {\cal R}_k \}$ are the ratios of CDS decay rates, 
\beq
{\cal R}_1 \equiv 
{\Gamma_{D^0 \to K^{*+}\pi^-} \over \Gamma_{D^0 \to K^{*0}\pi^0}} \ , 
\qquad {\cal R}_2 \equiv 
{\Gamma_{D^0 \to K^{*+}\pi^-} \over \Gamma_{D^+ \to K^{*0}\pi^+}} \ , 
\qquad {\cal R}_3 \equiv 
{\Gamma_{D^0 \to K^{*+}\pi^-} \over \Gamma_{D^+ \to K^{*+}\pi^0}} \ \ .
\label{a2}
\eeq

By eliminating $r$ and $\cos\bar\Delta$ from the relations in 
Eq.~(\ref{a1}), one obtains a cubic equation in the variable 
$R$.  However, there is an unphysical root $R = -1$, leaving 
the solution as a root of the quadratic equation
\beq
{\cal R}_2 {\cal R}_3 (2 {\cal R}_1 - 1) R^2 
+ \left(2 {\cal R}_2 {\cal R}_3 + {\cal R}_1 {\cal R}_3 
- 2 {\cal R}_1 {\cal R}_2 - {\cal R}_1 {\cal R}_2 {\cal R}_3 \right) R
+ {\cal R}_1 {\cal R}_2 - 2 {\cal R}_1 {\cal R}_3 = 0 \ \ .
\label{a3}
\eeq
It turns out that to obtain the physical solution it is 
necessary to choose the square root of the discriminant as 
{\it positive},   
\beq
R = {-b + \sqrt{b^2 - 4 a c} \over 2a} 
\label{a4}
\eeq
where 
\beqa
a &=& {\cal R}_2 {\cal R}_3 (2 {\cal R}_1 - 1) \ \ , 
\nonumber \\
b &=& 2 {\cal R}_2 {\cal R}_3 + {\cal R}_1 {\cal R}_3 
- 2 {\cal R}_1 {\cal R}_2 - {\cal R}_1 {\cal R}_2 {\cal R}_3 
\label{a5} \\
c &=& {\cal R}_1 {\cal R}_2 - 2 {\cal R}_1 {\cal R}_3 \ \ .
\nonumber
\eeqa
To see why, let us consider a hypothetical 
world with $c_+ = 0$.  Then since ${\cal H}_-^{\rm (CDS)}$ 
is an isoscalar operator, it follows that 
\beq
{\bar A}_3 = 0 \Longrightarrow r = 0 \qquad {\rm and} 
\qquad {\bar A}_b = {\bar A}_a \Longrightarrow R = 1 \ \ .
\label{a5a}
\eeq
As a consequence, we have 
\beq
{\cal R}_1 = 2\ ,\qquad {\cal R}_2 = 1\ , \qquad {\cal R}_3 = 2 \ \ ,
\label{a5b}
\eeq
from which the physical solution of Eq.~(\ref{a3}) is identified.
Returning to the real world, from $R$ one obtains $r$ 
\beq
r = \left[{ (2 - {\cal R}_1)(1 + R {\cal R}_2 ) - 2 (1 - {\cal R}_2 R^2)
(1 + {\cal R}_1) \over 2 (1 - {\cal R}_2 ) ( 1 + {\cal R}_1) 
- 2 (1 - 2 {\cal R}_1) ( 1 + R {\cal R}_2 )} \right]^{1/2}
\label{a6}
\eeq
and lastly $\cos\bar\Delta$,
\beq
\cos\bar\Delta = {2 - {\cal R}_1 + 2 r^2 (1 - 2 {\cal R}_1 ) 
\over 4 r (1 + {\cal R}_1 )} \ \ .
\label{a7}
\eeq

\subsubsection{Determining the Phase $\delta_{K^*\pi}^{\rm (ch)}$}
In the relation $\delta_{K^*\pi}^{\rm (ch)} \equiv
\delta_{-+}^{(K^*\pi)} - \delta_{+-}^{(K^*\pi)}$, 
the CF phase $\delta_{-+}^{(K^*\pi)}$ cannot be determined 
from the relations in Eq.~(\ref{cf2}) because one cannot know 
the individual values of $\delta_1$ and $\delta_3$.  However, 
the first relation in Eq.~(\ref{cf2}) does allow one to solve for 
$\delta_{-+}^{(K^*\pi)} - \delta_3$, 
\beq
\delta_{-+}^{(K^*\pi)} - \delta_3 = tan^{-1}\left[ \displaystyle{ 
\sin\left(\delta_1 - \delta_3\right) \over  
\cos\left(\delta_1 - \delta_3\right) + A_3/ 2 A_1  } \right] \ \ .
\label{cf5}
\eeq
Analogously, from the CDS relation in Eq.~(\ref{cds2}) we can solve 
for $\delta_{+-}^{(K^*\pi)} - \delta_3$, 
\beq
\delta_{+-}^{(K^*\pi)} - \bar\delta_3 = \tan^{-1} \left[{\sin\left(\bar\delta_1 
- \bar\delta_3\right) 
\over \cos\left(\bar\delta_1 - \bar\delta_3\right) - 
{\bar A}_3/ {\bar A}_a  }\right] \ \ , 
\label{d2}
\eeq
where $\delta_{+-}^{(K^*\pi)}$ is the phase of the CDS $D^0 \to K^{*+}\pi^-$ 
amplitude.  

Combining these two relations we find for $\delta_{K^*\pi}^{\rm (ch)}$ 
\beqa
\delta_{K^*\pi}^{\rm (ch)} &\equiv& \delta_{-+}^{(K^*\pi)} 
- \delta_{+-}^{(K^*\pi)} \nonumber \\
&=&  \delta_3 - {\bar\delta}_3   + tan^{-1}\left[ \displaystyle{ 
\sin\left(\delta_1 - \delta_3\right) \over  
\cos\left(\delta_1 - \delta_3\right) + A_3/ 2 A_1  } \right] 
- tan^{-1}\left[ \displaystyle{ 
\sin\left({\bar\delta}_1 - {\bar\delta}_3\right) \over  
 \cos\left({\bar\delta}_1 - {\bar\delta}_3\right) 
- {\bar A}_3/ {\bar A}_a  } \right] \ .
\label{del1} 
\eeqa
Under the assumptions that only $K_{\rm S}$ data is used 
and that isospin is a valid symmtery, we 
conclude that Eq.~(\ref{del1}) will be the best one can do 
in a purely experimental determination of $\delta_{K^*\pi}^{\rm (ch)}$. 
An expression for $\delta_{K^*\pi}^{\rm (ch)}$ itself is 
obtained only via dropping 
the contribution $\delta_3 - {\bar\delta}_3$.  
One might argue that these phases occur in 
exotic channels and should be individually small. 
Unlike the case of the $D \to {\bar K}\pi$ amplitudes, there 
is no SU(3) prediction that $\delta_{K^*\pi}^{\rm (ch)} = 0$. 

\subsection{Including $K_{\rm L}$ Data}
In the previous subsection, the avoidance of $K_{\rm L}$ data 
forced us to work with $D \to K^*\pi$ decays.  
The inclusion of $K_{\rm L}$ data allows us to return to 
the $D \to {\bar K}\pi$ and $D \to K\pi$ decays in the following.  
Each $D \to K_{\rm S,L} \pi$ mode will 
receive contributions from both CF and CDS sectors.
Writing the transition amplitudes for $D^0 \to K_{\rm S,L} \pi^0$ 
and $D^+ \to K_{\rm S,L} \pi^+$ in a generic notation, we have 
\beqa
{\cal M}_{D \to K_{\rm S} \pi} &=& {1 \over \sqrt{2}} \left[ 
|{\cal M}_{\rm CF}| e^{i \delta_{\rm CF}} -
|{\cal M}_{\rm CDS}| e^{i \delta_{\rm CDS}} \right] \ \ , 
\nonumber \\
{\cal M}_{D \to K_{\rm L} \pi} &=&  - {1 \over \sqrt{2}} \left[
|{\cal M}_{\rm CF}| e^{i \delta_{\rm CF}} + 
|{\cal M}_{\rm CDS}| e^{i \delta_{\rm CDS}} \right] \ \ . 
\label{L1}
\eeqa
The corresponding decay widths will each contain three terms,
\beqa
\Gamma_{D\to K_{\rm S} \pi} &=& {1\over 2} \Gamma_{\rm CF} 
- \sqrt{\Gamma_{\rm CF}} 
\sqrt{\Gamma_{\rm CDS}} \cos\left(\delta_{\rm CF} - 
\delta_{\rm CDS}\right) + {1\over 2} \Gamma_{\rm CDS} \ \ ,  
\nonumber \\
\Gamma_{D\to K_{\rm L} \pi} &=& {1\over 2} \Gamma_{\rm CF} 
+ \sqrt{\Gamma_{\rm CF}} 
\sqrt{\Gamma_{\rm CDS}} \cos\left(\delta_{\rm CF} - 
\delta_{\rm CDS}\right) + {1\over 2} \Gamma_{\rm CDS} \ \ .
\label{L2}
\eeqa
To our knowledge it has been standard in the PDG data compilation 
for $D \to {\bar K}^0 \pi$ to ignore all but the 
CF contribution by using the `factor of two 
rule' in Eq.~(\ref{int}) to infer the CF $D\to {\bar K}^0 X$ 
branching fraction from that of $D \to K_{\rm S} X$.  However, 
some account of sub-dominant terms is made in Ref.~\cite{cleo2} 
by attributing to their neglect a source of $10\%$ systematic error. 

In terms of Cabibbo counting, the contributions on the 
right hand side of Eq.~(\ref{L2}) go as 
$1:\theta_c^2:\theta_c^4$ or roughly $1:0.05:0.002$.  
Taking the sum of decay rates gives 
\beq
\Gamma_{D\to K_{\rm S} \pi} + \Gamma_{D\to K_{\rm L} \pi} = 
\Gamma_{\rm CF} + \Gamma_{\rm CDS} \ \ .
\label{L3}
\eeq
Since no existing facility can deliver $0.2\%$ sensitivity, 
the $\Gamma_{\rm CDS}$ contribution to this equation 
is negligible and one arrives at the kind of relation given 
earlier in Eq.~(\ref{int}). 

There {\it is}, however, the possibility of observing the 
${\cal O}(\theta_c^2)$ interference term via the asymmetry 
measurement~\cite{by,xing} 
\beq
{\cal A} \equiv {\Gamma_{D\to K_{\rm S} \pi} 
- \Gamma_{D\to K_{\rm L}\pi} \over 
\Gamma_{D\to K_{\rm S} \pi} + \Gamma_{D\to K_{\rm L} \pi} } 
\simeq - 2 { \sqrt{\Gamma_{\rm CDS}} \over 
\sqrt{\Gamma_{\rm CF}} } \cos\left(\delta_{\rm CF} - 
\delta_{\rm CDS}\right)  \ \ , 
\label{L4}
\eeq
or more specifically 
\beq
{\cal A}_{00} = - 2 \sqrt{\Gamma_{D^0\to K^0 \pi^0} \over 
\Gamma_{D^0\to {\bar K}^0 \pi^0} } 
\cos\left(\delta_{{\bar K}^0 \pi^0} - 
\delta_{K^0\pi^0}\right)  \ , \ \ 
{\cal A}_{0+} = - 2 \sqrt{\Gamma_{D^+\to K^0 \pi^+} \over 
\Gamma_{D^+\to {\bar K}^0 \pi^+} } 
\cos\left(\delta_{{\bar K}^0 \pi^+} - 
\delta_{K^0\pi^+}\right)  \ .
\label{asy}
\eeq
These asymmetries are ${\cal O}(\theta_c^2)$, so 
signals will occur at about the $5\%$ level.  Such 
measurements are difficult for existing B-factories but 
hopefully can be performed.

The detection of these asymmetries is clearly intriguing 
because they refer {\it directly} to $\delta_{\rm CF} - 
\delta_{\rm CDS}$.  Although 
the phase differences $\delta_{{\bar K}^0 \pi^0} - 
\delta_{K^0\pi^0}$ and $\delta_{{\bar K}^0 \pi^+} - 
\delta_{K^0\pi^+}$ in Eq.~(\ref{asy}) are for neutral modes (and 
not the charged case $\delta_{K^- \pi^+} - \delta_{K^+\pi^-}$)
it is nonetheless valuable information. At a rigorous level, 
it follows from the positivity of decay widths that a negative 
(positive) asymmetry would correspond a phase difference in the 
first (second) quadrant.  Beyond that one is forced into modelling 
$\Gamma_{\rm CDS}$. Since this contributes as a 
square root, the effect of model dependence is 
somewhat softened but still may be large.  We note that 
$\delta_{{\bar K}^0 \pi^0} - \delta_{K^0\pi^0} = 0$ in the 
SU(3) limit.

\vspace{0.2in}

We conclude this section by considering how to implement 
a complete data set for the $D \to K \pi$ decays.  It is 
understood from the preceding discussion that we organize 
the $K_{\rm S}$ and  $K_{\rm L}$ final states into sums and 
differences.  There will be a total of seven $D \to K \pi$ decays, 
of which three provide information on CF physics 
and four on CDS physics.  Defining $\bar\Gamma_k \equiv p 
{\bar A}_k^2/(8\pi m_D^2) \ (k = 1,3)$, we have for CF-related 
decays 
\begin{eqnarray}
\Gamma_{K^-\pi^+} &=& \Gamma_1 + \sqrt{\Gamma_1 \Gamma_3} 
\cos\left( \delta_1 - \delta_3 \right) + {1\over 4} \Gamma_3 
\nonumber \\
 \Gamma_{K_L\pi^0} + \Gamma_{K_S\pi^0} 
&=& {1 \over 2} \Gamma_1 - \sqrt{\Gamma_1 \Gamma_3} 
\cos\left( \delta_1 - \delta_3 \right) + {1\over 2} \Gamma_3 
+ \bar\Gamma_a + 4 \sqrt{\bar\Gamma_a \bar\Gamma_3} 
\cos\left( \bar\delta_1 - \bar\delta_3 \right) + 4 \bar\Gamma_3 
\nonumber \\
&\simeq& {1 \over 2} \Gamma_1 - \sqrt{\Gamma_1 \Gamma_3} 
\cos\left( \delta_1 - \delta_3 \right) + {1\over 2} \Gamma_3 
\nonumber \\
 \Gamma_{K_L\pi^+} + \Gamma_{K_S\pi^+} 
&=& {9\over 4} \Gamma_3 
+ 2 \bar\Gamma_b + 4 \sqrt{\bar\Gamma_b \bar\Gamma_3} 
\cos\left( \bar\delta_1 - \bar\delta_3 \right) + 2 \bar\Gamma_3 
\simeq {9\over 4} \Gamma_3  \ \ .
\label{dataCF}
\end{eqnarray}
In the latter two relations, we have made the approximation 
of discarding ${\cal O}(\theta_c^4)$ contributions 
(in accordance with the discussion around Eq.~(\ref{L3})).  
The approximate relations are seen to reproduce the content of 
Eq.~(\ref{cf4}).  For the CDS-related decays we have 
\begin{eqnarray}
\Gamma_{K^+\pi^-} &=& 2 \bar\Gamma_a - 4  \sqrt{\bar\Gamma_a \bar\Gamma_3} 
\cos\left( \bar\delta_1 - \bar\delta_3 \right) + 2 \bar\Gamma_3 
\nonumber \\
\Gamma_{K^+\pi^0} &=& \bar\Gamma_b - 4  \sqrt{\bar\Gamma_b \bar\Gamma_3} 
\cos\left( \bar\delta_1 - \bar\delta_3 \right) + 4 \bar\Gamma_3 
\nonumber \\
 \Gamma_{K_L\pi^0} - \Gamma_{K_S\pi^0} 
&=& - \sqrt{2} \sqrt{\Gamma_1 \bar\Gamma_a} 
\cos\left( \delta_1 - \bar\delta_1 \right) 
- 2 \sqrt{2} \sqrt{\Gamma_1 \bar\Gamma_3} 
\cos\left( \delta_1 - \bar\delta_3 \right) 
\nonumber \\
& & + \sqrt{2} \sqrt{\Gamma_3 \bar\Gamma_a} 
\cos\left( \delta_3 - \bar\delta_1 \right) 
+ 2 \sqrt{2} \sqrt{\Gamma_3 \bar\Gamma_3} 
\cos\left( \delta_3 - \bar\delta_3 \right) 
\nonumber \\
 \Gamma_{K_L\pi^+} - \Gamma_{K_S\pi^+} 
&=& 3 \sqrt{2} \sqrt{\Gamma_3 \bar\Gamma_b} 
\cos\left( \delta_3 - \bar\delta_1 \right) + 
3 \sqrt{2} \sqrt{\Gamma_3 \bar\Gamma_3} 
\cos\left( \delta_3 - \bar\delta_3 \right) \ \ ,
\label{dataCDS}
\end{eqnarray}
where $\bar\Gamma_k \equiv p {\bar A}_k^2/(8\pi m_D^2) \ 
(k = a,b,3)$.  Each term in the first two relations is 
${\cal O}(\theta_c^4)$ while each term in the latter two 
are ${\cal O}(\theta_c^2)$.  As an aid to analyzing 
these equations, we propose a simplified scenario with 
$\delta_3 = {\bar\delta}_3 = 0$ and 
$\delta_1 = \pi/2$. The approximate equations which result are 
\beqa
& & \Gamma_{K^+\pi^-} \simeq 2 \bar\Gamma_a - 4 
\sqrt{\bar\Gamma_a \bar\Gamma_3} \cos{\bar\delta}_1 
+ 2 \bar\Gamma_3  
\nonumber \\
& & \Gamma_{K^+\pi^0} \simeq  \bar\Gamma_b - 4 
\sqrt{\bar\Gamma_b \bar\Gamma_3} \cos{\bar\delta}_1 
+ 4 \bar\Gamma_3  
\nonumber \\
& & \Gamma_{K_{\rm L} \pi^0} - 
\Gamma_{K_{\rm S} \pi^0} \simeq - \sqrt{2} 
\sqrt{ \Gamma_1 {\bar \Gamma}_a}
\sin{\bar\delta}_1 + \sqrt{2} \sqrt{ \Gamma_3 {\bar \Gamma}_a}
\cos{\bar\delta}_1 + 2 \sqrt{2} \sqrt{ \Gamma_3 {\bar \Gamma}_3}
\nonumber \\
& & \Gamma_{K_{\rm L} \pi^+} - 
\Gamma_{K_{\rm S} \pi^+} \simeq 
3\sqrt{2} \sqrt{ \Gamma_3 \bar\Gamma_b  } \cos{\bar\delta}_1 + 
3\sqrt{2} \sqrt{ \Gamma_3 \bar\Gamma_3 } \ \ .
\label{L5}
\eeqa
In general, one must solve numerically for the unknowns 
$\bar\Gamma_a$, $\bar\Gamma_b$, $\Gamma_3$ and 
$\sin{\bar\delta}_1$.  Let us point out, however, the 
qualitative difference between the limiting cases ${\bar\delta}_1 
\simeq \pi/2 $ or ${\bar\delta}_1 \simeq 0$.  To study this 
difference, it is {\it not} enough to measure just the 
$K^+\pi^-$ and $K^+\pi^0$ final states; the $K_{\rm L,S} \pi$ 
modes are required as well.  It suffices to note here 
for ${\bar\delta}_1 \simeq \pi/2 $ that 
$(\Gamma_{K_{\rm L} \pi^0} - \Gamma_{K_{\rm S} \pi^0})^2/
\Gamma_{K^+\pi^-}\to \Gamma_1$ and  
$(\Gamma_{K_{\rm L} \pi^+} - \Gamma_{K_{\rm S} \pi^+})^2/
\Gamma_{K^+\pi^0}\to \Gamma_3\bar\Gamma_3/\bar\Gamma_b$,  
whereas for ${\bar\delta}_1 \simeq 0$ both ratios become 
$\Gamma_3$.  As ${\bar\delta}_1$ proceeds from $\pi/2$ 
to $0$, the first ratio decreases but the second 
increases by almost an order of magnitude.

\section{Conclusions}
The recent FOCUS experiment on $\Delta \Gamma_{\rm D}$ 
has yielded a signal at the several per cent level.  By comparison, 
this experimental result is over an order-of-magnitude larger 
than the value $y_{\rm cp} \simeq 0.8 \cdot 10^{-3}$ 
obtained in a theoretical analysis~\cite{blp} based 
on a sum over many $D^0$ decay modes.  In this paper,  
we have avoided the temptation to provide a theoretical prediction 
of our own for $\Delta\Gamma_{\rm D}$.  As stated earlier, 
we are not aware of any analytic approach in the charm region for which 
theoretical errors/uncertainties can be controlled.  We therefore feel 
that whether or not the FOCUS result holds up over time is for future 
{\it experimental} work to decide.  

At the very least, however, the E791, FOCUS, BELLE and CLEO studies 
serve to stimulate fresh thinking on a subject ($D^0$ mixing) 
that has long resisted progress. Our work in this paper 
has been to suggest further experimental work which would be of value:
\begin{enumerate}
\item We have described in Sect.~II both positive and negative 
aspects of various $D^0$ decays beyond those used in the 
E791, FOCUS and BELLE experiments.  In particular, we recommend that the 
$K_{\rm S}\phi$, $K_{\rm S}\omega$ and $K_{\rm S}\rho^0$ modes 
be given serious attention.  Each of these lies within the 
${\it CP} = -1$ sector, which heretofore has only been probed indirectly 
via the mixed-CP case of the $D \to (K^-\pi^+ + K^+\pi^-)$ transition.
\item In Sect.~III we divided our discussion of the strong phase 
$\delta \equiv \delta_{K\pi}^{\rm (ch)}$ into two parts: 
\subitem Supposing that accurate data on $K_{\rm L}$ final 
states is not forthcoming, we concluded that it will not be possible to 
probe the phase $\delta_{K\pi}$ experimentally, but that the 
$\delta_{K^*\pi}$ decays {\it would} be accessible.  Thus, 
we propose that branching fractions for 
the four CDS decays $D^0 \to K^{*+}\pi^-, K^{*0}\pi^0$ and 
$D^+\to K^{*+}\pi^0, K^0 \pi^{*+}$ be studied.  At present, there is 
data only for the $D^+ \to K^{*0}\pi^+$ transition, with a stated 
uncertainty of about $44\%$. Although any CDS branching fraction 
will be very small, the availability of copious charm production 
at B-factories and hadron colliders allows for the study of 
this hidden corner of charm 
physics.\footnote{After this paper was completed, an 
announcement appeared of a new CLEO measurement,
$y_{\rm CP}= - (1.1 \pm 2.5 \pm 1.4)\%$.~\cite{cleo3} 
This is consistent with previous 
results ({\it cf} Eq.~(\ref{intro2})). They also report
a first measurement of the CDS mode $D^0 \to K^+ \pi^- \pi^0$ 
which is a start of the exploration of the $K^*\pi$ CDS modes.}

\subitem We explored the eventuality that accurate data on 
$K_{\rm L}\pi$ final states will also be gathered.  In principle, the 
asymmetries of Eq.~(\ref{asy}) would provide direct examples of 
CF - CDS phase differences, but are hindered by the dependence 
on CDS branching fractions.  A more ambitious program would be 
to collect the complete set of CDS $K\pi$ data displayed in 
Eq.~(\ref{L5}).  In principle, this would allow for a 
determination of $\delta_{+-}^{(K\pi)}$ like that given in  
Eq.~(\ref{del1}) for $K^*\pi$.  Finally, we note that 
although our approach in this paper has been limited to 
what can be learned from just decay rates, the study 
of Dalitz distributions in multibody final states offers 
a separate opportunity for attacking the `$\delta_{+-}^{(K\pi)}$ 
problem'. 
\end{enumerate}

\acknowledgments
The research described here was supported in part by the 
National Science Foundation and by the Department of Energy.
We thank Guy Blaylock, Tom Browder, John Donoghue and Harry 
Nelson for their helpful input and Jonathan Link 
for a careful reading of the paper.

\end{document}